\documentclass[3p]{elsarticle}
\usepackage{amssymb}
\usepackage{amsmath}
\usepackage{graphicx}
\usepackage{subfigure}
\usepackage{psfrag}
\usepackage{color}
\biboptions{comma,square}
\journal{Physica A}

\providecommand\matlabtextAA{\color[rgb]{0.000,0.000,0.000}\fontsize{5}{5}\selectfont\strut}%
\psfrag{0037}[cl][cl]{\matlabtextAA $\theta=90^\circ$}%
\psfrag{0038}[cl][cl]{\matlabtextAA $\theta=60^\circ$}%
\psfrag{0039}[cl][cl]{\matlabtextAA $\theta=0^\circ$}%
\providecommand\matlabtextAB{\color[rgb]{0.000,0.000,0.000}\fontsize{8}{8}\selectfont\strut}%
\psfrag{0040}[bc][bc]{\matlabtextAB $\log_{10}|g(z)|$}%
\psfrag{0041}[tc][tc]{\matlabtextAB $\log_{10}r$}%
\psfrag{0042}[bc][bc]{\matlabtextAB $\mathrm{Im}[g(z)]$}%
\psfrag{0043}[tc][tc]{\matlabtextAB $\log_{10}r$}%
\psfrag{0044}[bc][bc]{\matlabtextAB $\mathrm{Re}[g(z)]$}%
\psfrag{0045}[tc][tc]{\matlabtextAB $\log_{10}r$}%
\def\matlabfragNegXTick{\mathord{\makebox[0pt][r]{$-$}}}
\providecommand\matlabtextAC{\color[rgb]{0.000,0.000,0.000}\fontsize{5}{5}\selectfont\strut}%
\psfrag{0000}[ct][ct]{\matlabtextAC $\matlabfragNegXTick 2$}%
\psfrag{0001}[ct][ct]{\matlabtextAC $\matlabfragNegXTick 1$}%
\psfrag{0002}[ct][ct]{\matlabtextAC $0$}%
\psfrag{0003}[ct][ct]{\matlabtextAC $1$}%
\psfrag{0004}[ct][ct]{\matlabtextAC $2$}%
\psfrag{0005}[ct][ct]{\matlabtextAC $3$}%
\psfrag{0015}[ct][ct]{\matlabtextAC $\matlabfragNegXTick 2$}%
\psfrag{0016}[ct][ct]{\matlabtextAC $\matlabfragNegXTick 1$}%
\psfrag{0017}[ct][ct]{\matlabtextAC $0$}%
\psfrag{0018}[ct][ct]{\matlabtextAC $1$}%
\psfrag{0019}[ct][ct]{\matlabtextAC $2$}%
\psfrag{0020}[ct][ct]{\matlabtextAC $3$}%
\psfrag{0026}[ct][ct]{\matlabtextAC $\matlabfragNegXTick 2$}%
\psfrag{0027}[ct][ct]{\matlabtextAC $\matlabfragNegXTick 1$}%
\psfrag{0028}[ct][ct]{\matlabtextAC $0$}%
\psfrag{0029}[ct][ct]{\matlabtextAC $1$}%
\psfrag{0030}[ct][ct]{\matlabtextAC $2$}%
\psfrag{0031}[ct][ct]{\matlabtextAC $3$}%
\psfrag{0006}[rc][rc]{\matlabtextAC $-16$}%
\psfrag{0007}[rc][rc]{\matlabtextAC $-14$}%
\psfrag{0008}[rc][rc]{\matlabtextAC $-12$}%
\psfrag{0009}[rc][rc]{\matlabtextAC $-10$}%
\psfrag{0010}[rc][rc]{\matlabtextAC $-8$}%
\psfrag{0011}[rc][rc]{\matlabtextAC $-6$}%
\psfrag{0012}[rc][rc]{\matlabtextAC $-4$}%
\psfrag{0013}[rc][rc]{\matlabtextAC $-2$}%
\psfrag{0014}[rc][rc]{\matlabtextAC $0$}%
\psfrag{0021}[rc][rc]{\matlabtextAC $-1$}%
\psfrag{0022}[rc][rc]{\matlabtextAC $-0.5$}%
\psfrag{0023}[rc][rc]{\matlabtextAC $0$}%
\psfrag{0024}[rc][rc]{\matlabtextAC $0.5$}%
\psfrag{0025}[rc][rc]{\matlabtextAC $1$}%
\psfrag{0032}[rc][rc]{\matlabtextAC $-1$}%
\psfrag{0033}[rc][rc]{\matlabtextAC $-0.5$}%
\psfrag{0034}[rc][rc]{\matlabtextAC $0$}%
\psfrag{0035}[rc][rc]{\matlabtextAC $0.5$}%
\psfrag{0036}[rc][rc]{\matlabtextAC $1$}%
\providecommand\matlabtextAD{\color[rgb]{0.000,0.000,0.000}\fontsize{8}{8}\selectfont\strut}%
\psfrag{0074}[cl][cl]{\matlabtextAD $\mathrm{Im}(z)$}%
\psfrag{0075}[br][br]{\matlabtextAD $\mathrm{Re}(z)\mathrm{\quad}$}%
\psfrag{0076}[bc][bc]{\matlabtextAD $\mathrm{Im}[g(z)]$}%
\psfrag{0077}[cl][cl]{\matlabtextAD $\mathrm{Im}(z)$}%
\psfrag{0078}[br][br]{\matlabtextAD $\mathrm{Re}(z)\mathrm{\quad}$}%
\psfrag{0079}[bc][bc]{\matlabtextAD $\mathrm{Re}[g(z)]$}%
\def\matlabfragNegXTick{\mathord{\makebox[0pt][r]{$-$}}}
\providecommand\matlabtextAE{\color[rgb]{0.000,0.000,0.000}\fontsize{5}{5}\selectfont\strut}%
\psfrag{0046}[ct][ct]{\matlabtextAE $\matlabfragNegXTick 5$}%
\psfrag{0047}[ct][ct]{\matlabtextAE $0$}%
\psfrag{0048}[ct][ct]{\matlabtextAE $5$}%
\psfrag{0049}[ct][ct]{\matlabtextAE $10$}%
\psfrag{0060}[ct][ct]{\matlabtextAE $\matlabfragNegXTick 5$}%
\psfrag{0061}[ct][ct]{\matlabtextAE $0$}%
\psfrag{0062}[ct][ct]{\matlabtextAE $5$}%
\psfrag{0063}[ct][ct]{\matlabtextAE $10$}%
\psfrag{0050}[rc][rc]{\matlabtextAE $-10$}%
\psfrag{0051}[rc][rc]{\matlabtextAE $-5$}%
\psfrag{0052}[rc][rc]{\matlabtextAE $0$}%
\psfrag{0053}[rc][rc]{\matlabtextAE $5$}%
\psfrag{0054}[rc][rc]{\matlabtextAE $10$}%
\psfrag{0064}[rc][rc]{\matlabtextAE $-10$}%
\psfrag{0065}[rc][rc]{\matlabtextAE $-5$}%
\psfrag{0066}[rc][rc]{\matlabtextAE $0$}%
\psfrag{0067}[rc][rc]{\matlabtextAE $5$}%
\psfrag{0068}[rc][rc]{\matlabtextAE $10$}%
\psfrag{0055}[cr][cr]{\matlabtextAE $-1$}%
\psfrag{0056}[cr][cr]{\matlabtextAE $-0.5$}%
\psfrag{0057}[cr][cr]{\matlabtextAE $0$}%
\psfrag{0058}[cr][cr]{\matlabtextAE $0.5$}%
\psfrag{0059}[cr][cr]{\matlabtextAE $1$}%
\psfrag{0069}[cr][cr]{\matlabtextAE $-1$}%
\psfrag{0070}[cr][cr]{\matlabtextAE $-0.5$}%
\psfrag{0071}[cr][cr]{\matlabtextAE $0$}%
\psfrag{0072}[cr][cr]{\matlabtextAE $0.5$}%
\psfrag{0073}[cr][cr]{\matlabtextAE $1$}%
\providecommand\matlabtextAX{\color[rgb]{0.000,0.000,0.000}\fontsize{5}{5}\selectfont\strut}%
\psfrag{0232}[cl][cl]{\matlabtextAX $f=0.025$}%
\psfrag{0233}[cl][cl]{\matlabtextAX $f=0.05$}%
\psfrag{0234}[cl][cl]{\matlabtextAX $f=0.1$}%
\psfrag{0235}[cl][cl]{\matlabtextAX $f=0.25$}%
\psfrag{0236}[cl][cl]{\matlabtextAX $f=0.5$}%
\providecommand\matlabtextAY{\color[rgb]{0.000,0.000,0.000}\fontsize{8}{8}\selectfont\strut}%
\psfrag{0237}[bc][bc]{\matlabtextAY $p_s(w)$}%
\psfrag{0238}[tc][tc]{\matlabtextAY $w$}%
\def\matlabfragNegXTick{\mathord{\makebox[0pt][r]{$-$}}}
\providecommand\matlabtextAZ{\color[rgb]{0.000,0.000,0.000}\fontsize{5}{5}\selectfont\strut}%
\psfrag{0218}[ct][ct]{\matlabtextAZ $0$}%
\psfrag{0219}[ct][ct]{\matlabtextAZ $0.5$}%
\psfrag{0220}[ct][ct]{\matlabtextAZ $1$}%
\psfrag{0221}[ct][ct]{\matlabtextAZ $1.5$}%
\psfrag{0222}[ct][ct]{\matlabtextAZ $2$}%
\psfrag{0223}[ct][ct]{\matlabtextAZ $2.5$}%
\psfrag{0224}[ct][ct]{\matlabtextAZ $3$}%
\psfrag{0225}[rc][rc]{\matlabtextAZ $0$}%
\psfrag{0226}[rc][rc]{\matlabtextAZ $0.5$}%
\psfrag{0227}[rc][rc]{\matlabtextAZ $1$}%
\psfrag{0228}[rc][rc]{\matlabtextAZ $1.5$}%
\psfrag{0229}[rc][rc]{\matlabtextAZ $2$}%
\psfrag{0230}[rc][rc]{\matlabtextAZ $2.5$}%
\psfrag{0231}[rc][rc]{\matlabtextAZ $3$}%
\providecommand\matlabtextBA{\color[rgb]{0.000,0.000,0.000}\fontsize{5}{5}\selectfont\strut}%
\psfrag{0257}[cl][cl]{\matlabtextBA $f=0.025$}%
\psfrag{0258}[cl][cl]{\matlabtextBA $f=0.05$}%
\psfrag{0259}[cl][cl]{\matlabtextBA $f=0.1$}%
\psfrag{0260}[cl][cl]{\matlabtextBA $f=0.25$}%
\psfrag{0261}[cl][cl]{\matlabtextBA $f=0.5$}%
\providecommand\matlabtextBB{\color[rgb]{0.000,0.000,0.000}\fontsize{8}{8}\selectfont\strut}%
\psfrag{0262}[bc][bc]{\matlabtextBB $\log_{10}p_s(w)$}%
\psfrag{0263}[tc][tc]{\matlabtextBB $w$}%
\def\matlabfragNegXTick{\mathord{\makebox[0pt][r]{$-$}}}
\providecommand\matlabtextBC{\color[rgb]{0.000,0.000,0.000}\fontsize{5}{5}\selectfont\strut}%
\psfrag{0239}[ct][ct]{\matlabtextBC $0$}%
\psfrag{0240}[ct][ct]{\matlabtextBC $1$}%
\psfrag{0241}[ct][ct]{\matlabtextBC $2$}%
\psfrag{0242}[ct][ct]{\matlabtextBC $3$}%
\psfrag{0243}[ct][ct]{\matlabtextBC $4$}%
\psfrag{0244}[ct][ct]{\matlabtextBC $5$}%
\psfrag{0245}[ct][ct]{\matlabtextBC $6$}%
\psfrag{0246}[ct][ct]{\matlabtextBC $7$}%
\psfrag{0247}[ct][ct]{\matlabtextBC $8$}%
\psfrag{0248}[rc][rc]{\matlabtextBC $-7$}%
\psfrag{0249}[rc][rc]{\matlabtextBC $-6$}%
\psfrag{0250}[rc][rc]{\matlabtextBC $-5$}%
\psfrag{0251}[rc][rc]{\matlabtextBC $-4$}%
\psfrag{0252}[rc][rc]{\matlabtextBC $-3$}%
\psfrag{0253}[rc][rc]{\matlabtextBC $-2$}%
\psfrag{0254}[rc][rc]{\matlabtextBC $-1$}%
\psfrag{0255}[rc][rc]{\matlabtextBC $0$}%
\psfrag{0256}[rc][rc]{\matlabtextBC $1$}%
\providecommand\matlabtextBP{\color[rgb]{0.000,0.000,0.000}\fontsize{5}{5}\selectfont\strut}%
\psfrag{0369}[cl][cl]{\matlabtextBP $f=0.9$}%
\psfrag{0370}[cl][cl]{\matlabtextBP $f=0.8$}%
\psfrag{0371}[cl][cl]{\matlabtextBP $f=0.7$}%
\psfrag{0372}[cl][cl]{\matlabtextBP $f=0.6$}%
\psfrag{0373}[cl][cl]{\matlabtextBP $f=0.5$}%
\providecommand\matlabtextBQ{\color[rgb]{0.000,0.000,0.000}\fontsize{8}{8}\selectfont\strut}%
\psfrag{0374}[bc][bc]{\matlabtextBQ $\log_{10}p_s(w)$}%
\psfrag{0375}[tc][tc]{\matlabtextBQ $\log_{10}w$}%
\def\matlabfragNegXTick{\mathord{\makebox[0pt][r]{$-$}}}
\providecommand\matlabtextBR{\color[rgb]{0.000,0.000,0.000}\fontsize{5}{5}\selectfont\strut}%
\psfrag{0350}[ct][ct]{\matlabtextBR $\matlabfragNegXTick 3$}%
\psfrag{0351}[ct][ct]{\matlabtextBR $\matlabfragNegXTick 2.5$}%
\psfrag{0352}[ct][ct]{\matlabtextBR $\matlabfragNegXTick 2$}%
\psfrag{0353}[ct][ct]{\matlabtextBR $\matlabfragNegXTick 1.5$}%
\psfrag{0354}[ct][ct]{\matlabtextBR $\matlabfragNegXTick 1$}%
\psfrag{0355}[ct][ct]{\matlabtextBR $\matlabfragNegXTick 0.5$}%
\psfrag{0356}[ct][ct]{\matlabtextBR $0$}%
\psfrag{0357}[ct][ct]{\matlabtextBR $0.5$}%
\psfrag{0358}[ct][ct]{\matlabtextBR $1$}%
\psfrag{0359}[ct][ct]{\matlabtextBR $1.5$}%
\psfrag{0360}[ct][ct]{\matlabtextBR $2$}%
\psfrag{0361}[rc][rc]{\matlabtextBR $-5$}%
\psfrag{0362}[rc][rc]{\matlabtextBR $-4$}%
\psfrag{0363}[rc][rc]{\matlabtextBR $-3$}%
\psfrag{0364}[rc][rc]{\matlabtextBR $-2$}%
\psfrag{0365}[rc][rc]{\matlabtextBR $-1$}%
\psfrag{0366}[rc][rc]{\matlabtextBR $0$}%
\psfrag{0367}[rc][rc]{\matlabtextBR $1$}%
\psfrag{0368}[rc][rc]{\matlabtextBR $2$}%
\providecommand\matlabtextBS{\color[rgb]{0.000,0.000,0.000}\fontsize{5}{5}\selectfont\strut}%
\psfrag{0394}[cl][cl]{\matlabtextBS $f=0.9$}%
\psfrag{0395}[cl][cl]{\matlabtextBS $f=0.8$}%
\psfrag{0396}[cl][cl]{\matlabtextBS $f=0.7$}%
\psfrag{0397}[cl][cl]{\matlabtextBS $f=0.6$}%
\psfrag{0398}[cl][cl]{\matlabtextBS $f=0.5$}%
\providecommand\matlabtextBT{\color[rgb]{0.000,0.000,0.000}\fontsize{8}{8}\selectfont\strut}%
\psfrag{0399}[bc][bc]{\matlabtextBT $\log_{10}p_s(w)$}%
\psfrag{0400}[tc][tc]{\matlabtextBT $w$}%
\def\matlabfragNegXTick{\mathord{\makebox[0pt][r]{$-$}}}
\providecommand\matlabtextBU{\color[rgb]{0.000,0.000,0.000}\fontsize{5}{5}\selectfont\strut}%
\psfrag{0376}[ct][ct]{\matlabtextBU $0$}%
\psfrag{0377}[ct][ct]{\matlabtextBU $5$}%
\psfrag{0378}[ct][ct]{\matlabtextBU $10$}%
\psfrag{0379}[ct][ct]{\matlabtextBU $15$}%
\psfrag{0380}[ct][ct]{\matlabtextBU $20$}%
\psfrag{0381}[ct][ct]{\matlabtextBU $25$}%
\psfrag{0382}[ct][ct]{\matlabtextBU $30$}%
\psfrag{0383}[ct][ct]{\matlabtextBU $35$}%
\psfrag{0384}[ct][ct]{\matlabtextBU $40$}%
\psfrag{0385}[ct][ct]{\matlabtextBU $45$}%
\psfrag{0386}[ct][ct]{\matlabtextBU $50$}%
\psfrag{0387}[rc][rc]{\matlabtextBU $-5$}%
\psfrag{0388}[rc][rc]{\matlabtextBU $-4$}%
\psfrag{0389}[rc][rc]{\matlabtextBU $-3$}%
\psfrag{0390}[rc][rc]{\matlabtextBU $-2$}%
\psfrag{0391}[rc][rc]{\matlabtextBU $-1$}%
\psfrag{0392}[rc][rc]{\matlabtextBU $0$}%
\psfrag{0393}[rc][rc]{\matlabtextBU $1$}%
\providecommand\matlabtextCB{\color[rgb]{0.000,0.000,0.000}\fontsize{8}{8}\selectfont\strut}%
\psfrag{0483}[bc][bc]{\matlabtextCB $\log_{10}n(w)$}%
\psfrag{0484}[tc][tc]{\matlabtextCB $\log_{10}w\;(\mathrm{m.u.})$}%
\def\matlabfragNegXTick{\mathord{\makebox[0pt][r]{$-$}}}
\providecommand\matlabtextCC{\color[rgb]{0.000,0.000,0.000}\fontsize{5}{5}\selectfont\strut}%
\psfrag{0464}[ct][ct]{\matlabtextCC $\matlabfragNegXTick 1$}%
\psfrag{0465}[ct][ct]{\matlabtextCC $\matlabfragNegXTick 0.5$}%
\psfrag{0466}[ct][ct]{\matlabtextCC $0$}%
\psfrag{0467}[ct][ct]{\matlabtextCC $0.5$}%
\psfrag{0468}[ct][ct]{\matlabtextCC $1$}%
\psfrag{0469}[ct][ct]{\matlabtextCC $1.5$}%
\psfrag{0470}[ct][ct]{\matlabtextCC $2$}%
\psfrag{0471}[ct][ct]{\matlabtextCC $2.5$}%
\psfrag{0472}[ct][ct]{\matlabtextCC $3$}%
\psfrag{0473}[ct][ct]{\matlabtextCC $3.5$}%
\psfrag{0474}[ct][ct]{\matlabtextCC $4$}%
\psfrag{0475}[rc][rc]{\matlabtextCC $-1$}%
\psfrag{0476}[rc][rc]{\matlabtextCC $0$}%
\psfrag{0477}[rc][rc]{\matlabtextCC $1$}%
\psfrag{0478}[rc][rc]{\matlabtextCC $2$}%
\psfrag{0479}[rc][rc]{\matlabtextCC $3$}%
\psfrag{0480}[rc][rc]{\matlabtextCC $4$}%
\psfrag{0481}[rc][rc]{\matlabtextCC $5$}%
\psfrag{0482}[rc][rc]{\matlabtextCC $6$}%
\providecommand\matlabtextCD{\color[rgb]{0.000,0.000,0.000}\fontsize{8}{8}\selectfont\strut}%
\psfrag{0504}[bc][bc]{\matlabtextCD $\log_{10}n(w)$}%
\psfrag{0505}[tc][tc]{\matlabtextCD $w\;(\mathrm{m.u.})$}%
\def\matlabfragNegXTick{\mathord{\makebox[0pt][r]{$-$}}}
\providecommand\matlabtextCE{\color[rgb]{0.000,0.000,0.000}\fontsize{5}{5}\selectfont\strut}%
\psfrag{0485}[ct][ct]{\matlabtextCE $0$}%
\psfrag{0486}[ct][ct]{\matlabtextCE $100$}%
\psfrag{0487}[ct][ct]{\matlabtextCE $200$}%
\psfrag{0488}[ct][ct]{\matlabtextCE $300$}%
\psfrag{0489}[ct][ct]{\matlabtextCE $400$}%
\psfrag{0490}[ct][ct]{\matlabtextCE $500$}%
\psfrag{0491}[ct][ct]{\matlabtextCE $600$}%
\psfrag{0492}[ct][ct]{\matlabtextCE $700$}%
\psfrag{0493}[rc][rc]{\matlabtextCE $-1$}%
\psfrag{0494}[rc][rc]{\matlabtextCE $-0.5$}%
\psfrag{0495}[rc][rc]{\matlabtextCE $0$}%
\psfrag{0496}[rc][rc]{\matlabtextCE $0.5$}%
\psfrag{0497}[rc][rc]{\matlabtextCE $1$}%
\psfrag{0498}[rc][rc]{\matlabtextCE $1.5$}%
\psfrag{0499}[rc][rc]{\matlabtextCE $2$}%
\psfrag{0500}[rc][rc]{\matlabtextCE $2.5$}%
\psfrag{0501}[rc][rc]{\matlabtextCE $3$}%
\psfrag{0502}[rc][rc]{\matlabtextCE $3.5$}%
\psfrag{0503}[rc][rc]{\matlabtextCE $4$}%
\providecommand\matlabtextCM{\color[rgb]{0.000,0.000,0.000}\fontsize{8}{8}\selectfont\strut}%
\psfrag{0589}[bc][bc]{\matlabtextCM $S_s$}%
\psfrag{0590}[cc][cc]{\matlabtextCM $\sigma_s^2$}%
\def\matlabfragNegXTick{\mathord{\makebox[0pt][r]{$-$}}}
\providecommand\matlabtextCN{\color[rgb]{0.000,0.000,0.000}\fontsize{5}{5}\selectfont\strut}%
\psfrag{0574}[ct][ct]{\matlabtextCN $10^{-2}$}%
\psfrag{0575}[ct][ct]{\matlabtextCN $10^{-1}$}%
\psfrag{0576}[ct][ct]{\matlabtextCN $10^{0}$}%
\psfrag{0577}[ct][ct]{\matlabtextCN $10^{1}$}%
\psfrag{0578}[rc][rc]{\matlabtextCN $-1$}%
\psfrag{0579}[rc][rc]{\matlabtextCN $-0.8$}%
\psfrag{0580}[rc][rc]{\matlabtextCN $-0.6$}%
\psfrag{0581}[rc][rc]{\matlabtextCN $-0.4$}%
\psfrag{0582}[rc][rc]{\matlabtextCN $-0.2$}%
\psfrag{0583}[rc][rc]{\matlabtextCN $0$}%
\psfrag{0584}[rc][rc]{\matlabtextCN $0.2$}%
\psfrag{0585}[rc][rc]{\matlabtextCN $0.4$}%
\psfrag{0586}[rc][rc]{\matlabtextCN $0.6$}%
\psfrag{0587}[rc][rc]{\matlabtextCN $0.8$}%
\psfrag{0588}[rc][rc]{\matlabtextCN $1$}%
\providecommand\matlabtextCQ{\color[rgb]{0.000,0.000,0.000}\fontsize{8}{8}\selectfont\strut}%
\psfrag{0627}[bc][bc]{\matlabtextCQ $S(t)$}%
\psfrag{0628}[tc][tc]{\matlabtextCQ $t\;(\mathrm{steps})$}%
\def\matlabfragNegXTick{\mathord{\makebox[0pt][r]{$-$}}}
\providecommand\matlabtextCR{\color[rgb]{0.000,0.000,0.000}\fontsize{5}{5}\selectfont\strut}%
\psfrag{0610}[ct][ct]{\matlabtextCR $0$}%
\psfrag{0611}[ct][ct]{\matlabtextCR $10$}%
\psfrag{0612}[ct][ct]{\matlabtextCR $20$}%
\psfrag{0613}[ct][ct]{\matlabtextCR $30$}%
\psfrag{0614}[ct][ct]{\matlabtextCR $40$}%
\psfrag{0615}[ct][ct]{\matlabtextCR $50$}%
\psfrag{0616}[ct][ct]{\matlabtextCR $60$}%
\psfrag{0617}[ct][ct]{\matlabtextCR $70$}%
\psfrag{0618}[ct][ct]{\matlabtextCR $80$}%
\psfrag{0619}[ct][ct]{\matlabtextCR $90$}%
\psfrag{0620}[ct][ct]{\matlabtextCR $100$}%
\psfrag{0621}[rc][rc]{\matlabtextCR $0$}%
\psfrag{0622}[rc][rc]{\matlabtextCR $0.05$}%
\psfrag{0623}[rc][rc]{\matlabtextCR $0.1$}%
\psfrag{0624}[rc][rc]{\matlabtextCR $0.15$}%
\psfrag{0625}[rc][rc]{\matlabtextCR $0.2$}%
\psfrag{0626}[rc][rc]{\matlabtextCR $0.25$}%
\providecommand\matlabtextCW{\color[rgb]{0.000,0.000,0.000}\fontsize{8}{8}\selectfont\strut}%
\psfrag{0683}[bc][bc]{\matlabtextCW $\log_{10}\xi(w)$}%
\psfrag{0684}[tc][tc]{\matlabtextCW $w$}%
\def\matlabfragNegXTick{\mathord{\makebox[0pt][r]{$-$}}}
\providecommand\matlabtextCX{\color[rgb]{0.000,0.000,0.000}\fontsize{5}{5}\selectfont\strut}%
\psfrag{0669}[ct][ct]{\matlabtextCX $0$}%
\psfrag{0670}[ct][ct]{\matlabtextCX $0.5$}%
\psfrag{0671}[ct][ct]{\matlabtextCX $1$}%
\psfrag{0672}[ct][ct]{\matlabtextCX $1.5$}%
\psfrag{0673}[ct][ct]{\matlabtextCX $2$}%
\psfrag{0674}[ct][ct]{\matlabtextCX $2.5$}%
\psfrag{0675}[ct][ct]{\matlabtextCX $3$}%
\psfrag{0676}[rc][rc]{\matlabtextCX $-5$}%
\psfrag{0677}[rc][rc]{\matlabtextCX $-4$}%
\psfrag{0678}[rc][rc]{\matlabtextCX $-3$}%
\psfrag{0679}[rc][rc]{\matlabtextCX $-2$}%
\psfrag{0680}[rc][rc]{\matlabtextCX $-1$}%
\psfrag{0681}[rc][rc]{\matlabtextCX $0$}%
\psfrag{0682}[rc][rc]{\matlabtextCX $1$}%
\providecommand\matlabtextCY{\color[rgb]{0.000,0.000,0.000}\fontsize{8}{8}\selectfont\strut}%
\psfrag{0707}[bc][bc]{\matlabtextCY $w_i$}%
\psfrag{0708}[tc][tc]{\matlabtextCY $i$}%
\def\matlabfragNegXTick{\mathord{\makebox[0pt][r]{$-$}}}
\providecommand\matlabtextCZ{\color[rgb]{0.000,0.000,0.000}\fontsize{5}{5}\selectfont\strut}%
\psfrag{0685}[ct][ct]{\matlabtextCZ $0$}%
\psfrag{0686}[ct][ct]{\matlabtextCZ $100$}%
\psfrag{0687}[ct][ct]{\matlabtextCZ $200$}%
\psfrag{0688}[ct][ct]{\matlabtextCZ $300$}%
\psfrag{0689}[ct][ct]{\matlabtextCZ $400$}%
\psfrag{0690}[ct][ct]{\matlabtextCZ $500$}%
\psfrag{0691}[ct][ct]{\matlabtextCZ $600$}%
\psfrag{0692}[ct][ct]{\matlabtextCZ $700$}%
\psfrag{0693}[ct][ct]{\matlabtextCZ $800$}%
\psfrag{0694}[ct][ct]{\matlabtextCZ $900$}%
\psfrag{0695}[ct][ct]{\matlabtextCZ $1000$}%
\psfrag{0696}[rc][rc]{\matlabtextCZ $0$}%
\psfrag{0697}[rc][rc]{\matlabtextCZ $0.2$}%
\psfrag{0698}[rc][rc]{\matlabtextCZ $0.4$}%
\psfrag{0699}[rc][rc]{\matlabtextCZ $0.6$}%
\psfrag{0700}[rc][rc]{\matlabtextCZ $0.8$}%
\psfrag{0701}[rc][rc]{\matlabtextCZ $1$}%
\psfrag{0702}[rc][rc]{\matlabtextCZ $1.2$}%
\psfrag{0703}[rc][rc]{\matlabtextCZ $1.4$}%
\psfrag{0704}[rc][rc]{\matlabtextCZ $1.6$}%
\psfrag{0705}[rc][rc]{\matlabtextCZ $1.8$}%
\psfrag{0706}[rc][rc]{\matlabtextCZ $2$}%
\providecommand\matlabtextDC{\color[rgb]{0.000,0.000,0.000}\fontsize{8}{8}\selectfont\strut}%
\psfrag{0740}[bc][bc]{\matlabtextDC $G_s$}%
\psfrag{0741}[cc][cc]{\matlabtextDC $\sigma_s^2$}%
\def\matlabfragNegXTick{\mathord{\makebox[0pt][r]{$-$}}}
\providecommand\matlabtextDD{\color[rgb]{0.000,0.000,0.000}\fontsize{5}{5}\selectfont\strut}%
\psfrag{0725}[ct][ct]{\matlabtextDD $10^{-2}$}%
\psfrag{0726}[ct][ct]{\matlabtextDD $10^{-1}$}%
\psfrag{0727}[ct][ct]{\matlabtextDD $10^{0}$}%
\psfrag{0728}[ct][ct]{\matlabtextDD $10^{1}$}%
\psfrag{0729}[rc][rc]{\matlabtextDD $0$}%
\psfrag{0730}[rc][rc]{\matlabtextDD $0.1$}%
\psfrag{0731}[rc][rc]{\matlabtextDD $0.2$}%
\psfrag{0732}[rc][rc]{\matlabtextDD $0.3$}%
\psfrag{0733}[rc][rc]{\matlabtextDD $0.4$}%
\psfrag{0734}[rc][rc]{\matlabtextDD $0.5$}%
\psfrag{0735}[rc][rc]{\matlabtextDD $0.6$}%
\psfrag{0736}[rc][rc]{\matlabtextDD $0.7$}%
\psfrag{0737}[rc][rc]{\matlabtextDD $0.8$}%
\psfrag{0738}[rc][rc]{\matlabtextDD $0.9$}%
\psfrag{0739}[rc][rc]{\matlabtextDD $1$}%
\providecommand{\norm}[1]{\lVert#1\rVert}

\begin{document}

\begin{frontmatter}

\title{Laplace transform analysis of a multiplicative asset transfer model}

\author[unimelb]{Andrey Sokolov}
\author[unimelb]{Andrew Melatos}
\author[unimelb,portland]{Tien Kieu}

\address[unimelb]{School of Physics, University of Melbourne, Parkville, VIC 3010, Australia}
\address[portland]{Centre for Atom Optics and Ultrafast Spectroscopy, Swinburne University of Technology, Hawthorn, VIC 3122, Australia}

\begin{abstract}
We analyze a simple asset transfer model in which the transfer amount is a fixed fraction $f$ of the giver's wealth.
The model is analyzed in a new way by Laplace transforming the master equation, 
solving it analytically and numerically for the steady-state distribution,
and exploring the solutions for various values of $f\in(0,1)$.
The Laplace transform analysis is superior to agent-based simulations as it does not depend on the number of agents,
enabling us to study entropy and inequality in regimes that are costly to address with simulations.
We demonstrate that Boltzmann entropy is not a suitable (e.g.~non-monotonic) measure of disorder in a multiplicative asset transfer system
and suggest an asymmetric stochastic process that is equivalent to the asset transfer model.
\end{abstract}


\end{frontmatter}

\section{Introduction}
\label{section.intro}
A vibrant research theme in econophysics is the analysis of asset exchange models.
In these models, a large number of agents iteratively exchange  assets,
typically representing monetary amounts.
In the simplest model that has been considered, the transfer amount is constant
and independent of the agent's wealth, producing an exponential wealth distribution in the steady state (see \cite{yakovenko2009colloquium} for a review).
More complicated fractional exchange models
have also been considered by several authors \cite{chatterjee2007kinetic,hayes2002follow}, in which the size of each transfer
is a linear function of the wealths of the agents involved in the exchange.

It has been found both analytically and numerically that the steady-state wealth probability distribution function $p_s(w)$
in fractional exchange models depends strongly on the parameters that characterize the exchange 
\cite{matthes2008steady}.
Certain parameter values or exchange rules yield a strongly peaked distribution with an exponential tail,
while other values yield a broad distribution with  Pareto-like qualities.
The dichotomy is exemplified by two simple models.
If the transfer amount is a fixed fraction $f$ of the giver's wealth 
(the giver is the agent who surrenders the asset in the transfer), 
then the resulting steady-state distribution is strongly peaked and decays exponentially in the tail.
If, on the other hand, the transfer amount is a fixed fraction $f$ of the poorer agent's wealth, 
then one finds a broad 
steady-state distribution, which can be fitted well by a power law with exponent $-1$
across a broad interval of wealths.
In this paper, we refer to these two models as the giver scheme and the poorer scheme 
respectively.\footnote{They are also known as the theft-and-fraud and yard-sale models respectively (see \cite{hayes2002follow}).}

In some of the asset exchange models considered in \cite{chakraborti2000statistical} and several other studies
\cite{chatterjee2007kinetic,chatterjee2005master},
the fractional exchange amount is a random linear combination of the wealths of the participating agents.
The controlling parameter is the saving propensity, $\lambda$, which
determines the fraction of the agents' wealths that they do not offer to exchange.
Comparing the output of simulations for the giver scheme and the exchange schemes based on the saving propensity,
one observes 
that the schemes are closely related, with $f\approx0$ corresponding to $\lambda\approx1$.
If the saving propensity is the same for all the agents, then
the resulting steady-state distributions are similar to those obtained for the giver scheme.\
On the other hand, if the saving propensity is uniformly distributed, the steady-state distribution
is a power law, $p_s(w)\propto w ^{-2}$.
In a recent study based on numerical simulations \cite{AliSaif2007448}, it was found
that a combination of the poorer and giver schemes in one simulation results in a power-law wealth
distribution whose exponent depends on the relative contributions of the two schemes.
The more agents follow the giver scheme, the greater the exponent.

Asset exchange models can be treated analytically via a kinetic or master equation, 
which tracks the rate of change of the number of agents at any given wealth.
In particular, the master equation for the giver scheme has been derived in \cite{ispolatov1998wealth}.
These authors found an expression for the second moment in the steady state,
which agrees with the expression
found in \cite{angle2006inequality} for a similar model by assuming the gamma distribution of wealth.
The standard deviation  converges to its steady-state value exponentially on a time scale $\sim [f(1-f)]^{-1}$.
The authors also found the asymptotic behaviour of the wealth distribution at small values of wealth.
The master equation for the poorer scheme was derived recently \cite{moukarzel2007wealth},
but its solutions have not yet been studied.
In \cite{chatterjee2005master}, the authors derived the kinetic equation for the case of uniformly distributed 
saving propensity. 
They demonstrated that the solution follows a power law with the same exponent as in the simulations.
The kinetic equation approach was also used in \cite{slanina2004inelastically} to analyze self-similar  solutions
of a non-conservative asset exchange system.
The author found a closed-form solution in the limit of continuous trading by means of the Laplace transform
and observed that the distribution exhibits power-law behaviour at large wealths.

The dependence of the relaxation time on the exchange parameters has been investigated numerically 
in \cite{patriarca2007relaxation} for the models considered in \cite{chatterjee2007kinetic,chatterjee2005master}.
The relaxation time-scale was found numerically to scale as $\sim(1-\lambda)^{-1}$.
This is consistent with the values found analytically in the giver scheme for the standard deviation. 
The authors also considered how the relaxation time depends on the number of agents
but failed to find any significant trend.

Recently, much effort has been directed profitably at developing more sophisticated and realistic multi-agent models
to be analyzed by means of numerical simulations.
In the present paper we take the opposite tack and return instead to one of the simplest multiplicative models, the giver scheme.
We show that its master equation can be solved efficiently by a Laplace transform technique.
Armed with this new tool backed by multi-agent simulations, we identify the following new properties of the system.
(1)~We get precise values of various quantities such as the steady-state entropy
as a function of the model parameter $f$, independently of the number of agents.
(2)~We explore the thinly studied regime $1/2< f<1$ and identify its unusual properties, e.g. oscillations in $p_s(w)$.
(3)~Using multi-agent simulations, 
we investigate how the Boltzmann entropy evolves with time as the system approaches equilibrium
and argue that the Boltzmann entropy is not a suitable entropy for the giver scheme,
even though the system is closed and conservative.
(4)~We propose a simple asymmetric stochastic process that is equivalent to the giver scheme.
(5)~We investigate how the degree of inequality, characterized by the Gini coefficient, depends on $f$.
(6)~Finally, we apply phase-space techniques from statistical mechanics to the giver scheme 
in order to illuminate the difficulties and opportunities that this asset transfer model presents.

\section{Giver scheme}
\label{section.model}

We consider a simple asset transfer model,
in which the transfer amount 
is equal to a fixed fraction of the giver's wealth.\footnote{The giver is also called the payer or the loser in the literature.}
If $w_g$ is the giver's wealth and $w_r$ is the receiver's wealth prior to the transfer, then
their wealths after the transfer are given by $w_g-\Delta w$ and $w_r+\Delta w$ respectively,
with $\Delta w=fw_g$ and  $f\in(0,1)$. 
The model comprises a large number of agents,
who are assigned wealths initially according to some distribution.
The transfers are assumed to take place over a fixed time interval $\Delta t$.
At each discrete time $t_i$, the agents are divided randomly into pairs and 
the transfer formula is applied to each pair.
The transfers are complete by the time $t_{i+1}=t_i+\Delta t$ and the process repeats at the time $t_{i+1}$.
The probability of drawing any pair is the same.
In each pair, the giver is assigned randomly regardless of the wealths of the agents.

The master equation for this system was derived in \cite{ispolatov1998wealth} and is given by
\begin{equation}\label{master}
\frac{\partial p(w,t)}{\partial t}
=
-
p(w,t)
+
\frac{1}{2(1-f)}
p\left(\frac{w}{1-f},t\right)
+\frac{1}{2f}
\int_0^w
dw'\;
p
\left(
\frac{w-w'}{f}
,t
\right)
p(w',t)
.
\end{equation}
It is easy to verify that the mean of the distribution, $\mu_1=\int_{0}^{\infty}dw\;wp(w,t)$, does not depend on time.
Upon integrating by parts, one arrives at the evolution equation
\begin{equation}\label{mu2t}
\frac{d\mu_2(t)}{dt}
=
-f(1-f)\mu_2 +f\mu_1^2
\end{equation}
for the second moment, $\mu_2(t)=\int_{0}^{\infty}dw\;w^2p(w,t)$, first reported in \cite{ispolatov1998wealth}.
This equation can be solved for the variance
\begin{equation}\label{std-time}
\sigma^2(t)
=
\mu_2(t)-\mu_1^2
=
\left(
  \mu_2(0)
  -
  \frac{\mu_1^2}{1-f}
\right)
e^{-f(1-f)t} 
+
\frac{f\mu_1^2}{1-f}
.
\end{equation}
In the steady state, one has $\sigma_s=\sigma(t\to\infty)=\mu_1 [f/(1-f)]^{1/2}$.
For simplicity, we assume henceforth that 
the mean of $p(w,t)$ equals unity.\footnote{If the mean $\mu_1\ne1$,
one can consider the function $q(x)=\mu_1 p(\mu_1 x)$, which is normalized and has unit mean.} 
In the following sections, we are mostly concerned with the steady-state distribution $p_s(w)=p(w,t\to\infty)$.

\section{Laplace transform of the master equation}
\label{section.master}
The Laplace transform of the master equation in the steady state is given by
\begin{equation}\label{fe}
g(z)
=
\frac{1}{2}
g(z-fz)
+
\frac{1}{2}
g(z)g(fz),
\end{equation}
with $g(z)=\int_{0}^{\infty}dw\; e^{-zw} p_s(w)$.
Note that the functional equation~\eqref{fe} applies to
any integral transform whose kernel depends only on the product of the arguments of the function and its transform.
For $f=1/2$, the functional equation has a closed form solution
\begin{equation}
g(z)
=
\frac{1}{1+Cz},
\end{equation}
where $C$ is a complex-valued constant.
Using the definition of the transform, we have $g(0)=1$ from the normalization of $p_s(w)$ and 
$g'(0)=-1$ from the assumption that $p_s(w)$ has unit mean, 
which gives $C=1$.
Applying the inverse Laplace transform to this solution gives the exponential distribution, 
$p_s(w)=e^{-w}$, which was obtained in \cite{ispolatov1998wealth} by substituting simple ``test'' functions into the master equation.
No closed-form solutions have been found for other values of $f\in(0,1)$.

The Taylor expansion of $g(z)$ at $z=0$ can be derived by substituting the expansion in \eqref{fe} 
and using $g(0)=1$ and $g'(0)=-1$.
For the first four terms of the expansion, this procedure gives
\begin{equation}\label{g0}
g(z)=1-z+\frac{1}{2(1-f)}z^2 - \frac{1+f}{6(1-f)^2}z^3+O(z^4).
\end{equation}
In general, for $g(z)=\sum_{n=0}^{\infty}a_n(-z)^n/n!$, $a_0=1$, and $a_1=1$, we obtain 
\begin{equation}\label{recursive}
a_n=
\sum_{k=1}^{n-1} 
\binom{n}{k}%{n\choose k}
\frac{f^k a_k a_{n-k}}{1-f^n-(1-f)^n}
\quad
\textrm{for } n>1.
\end{equation}
Since $g(-z)$ is the moment-generating function for the distribution $p_s(w)$,
the $n$-th moment of the distribution $\mu_n$ equal $a_n$, i.e.\ all moments 
of the steady-state wealth distribution can be computed for any $f$ 
using the recursive formula~\eqref{recursive}.

Using the Taylor expansion, one has $a_n\to1$ as $f\to0$ and hence $g(z)\to e^{-z}$.
Note that the functional equation~\eqref{fe} becomes an identity for $f=0$ and $g(0)=1$.
Taking the inverse Laplace transform of $g(z)=e^{-z}$, formally one gets $p_s(w)=\delta(w-1)$.
However, this wealth distribution is never reached
because the relaxation time scale $t_r=[f(1-f)]^{-1}$ determined from~\eqref{std-time} tends to infinity in this limit.
Indeed, $p_s(w)$ is equal to the initial distribution if $f=0$.
On the other hand, the Taylor expansion does not have a limit as $f\to1$.
The functional equation~\eqref{fe} has the solution $g(z)=1$ when $f=1$, 
but it does not satisfy the condition $g'(0)=-1$.
It appears that $p_s(w)$ does not have a proper limit as $f\to1$.
Note also that the relaxation time tends to infinity as $f$ approaches unity as well.

The asymptotic behaviour of $g(z)$ at infinity can also be deduced readily from the functional equation.
In the directions in the complex plane for which one has $g(z)\to0$ as $|z|\to\infty$,
the equation 
\begin{equation}\label{fez infinity}
g(z-fz)=2 g(z)
\end{equation}
must be approximately true for large enough $|z|$.
Assuming a power-law shape $|g(z)|\propto|z|^{-\alpha}$ as $|z|\to\infty$,
equation~\eqref{fez infinity} gives 
\begin{equation}\label{alpha}
\alpha=\frac{-1}{\log_2(1-f)}.
\end{equation}
By Watson's lemma \cite{davies2002integral}, this is consistent with the asymptotic behaviour $p(w)\propto w^{\alpha-1}$ as $w\to0$
that was found in~\cite{ispolatov1998wealth} by the method of dominant balance. 

The functional equation~\eqref{fe} can be solved iteratively when it is cast in the form
\begin{equation}\label{iteration equation}
g_{i+1}(z)
=
\frac{g_i(z-fz)}{2-g_i(fz)},
\end{equation}
where $g_i(z)$ is the $i$-th iteration.
Experimentation shows that the choice  $g_0(z)=1/(1+z)$ works well for all $f$.
A detailed description of the computational procedure is given in~\ref{appendix.iterations};
there are some subtleties involved in the choice of grid and interpolation method.
\begin{figure}[t!]
{\centering
\subfigure[]{\includegraphics{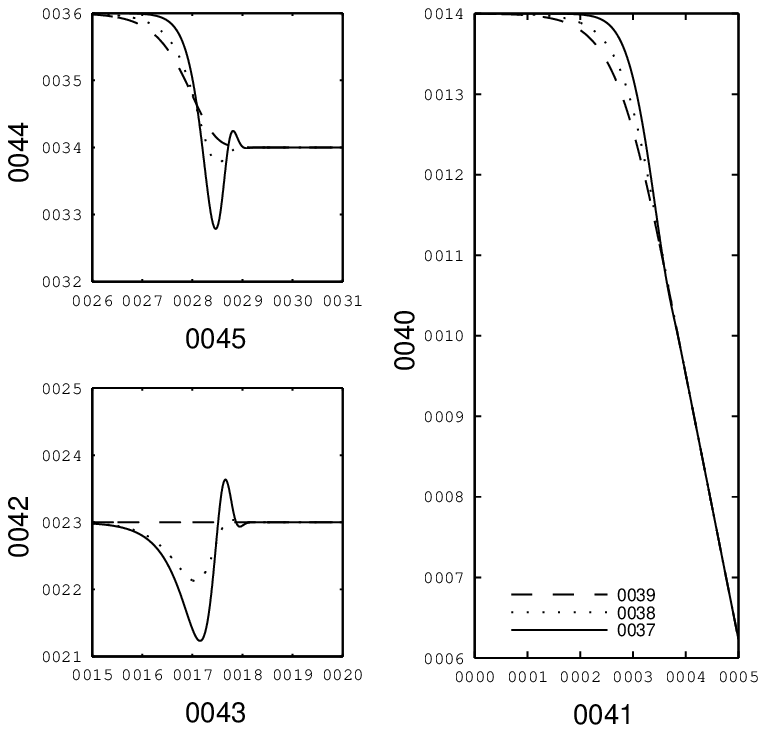} \label{gz0.1}}
\subfigure[]{\includegraphics{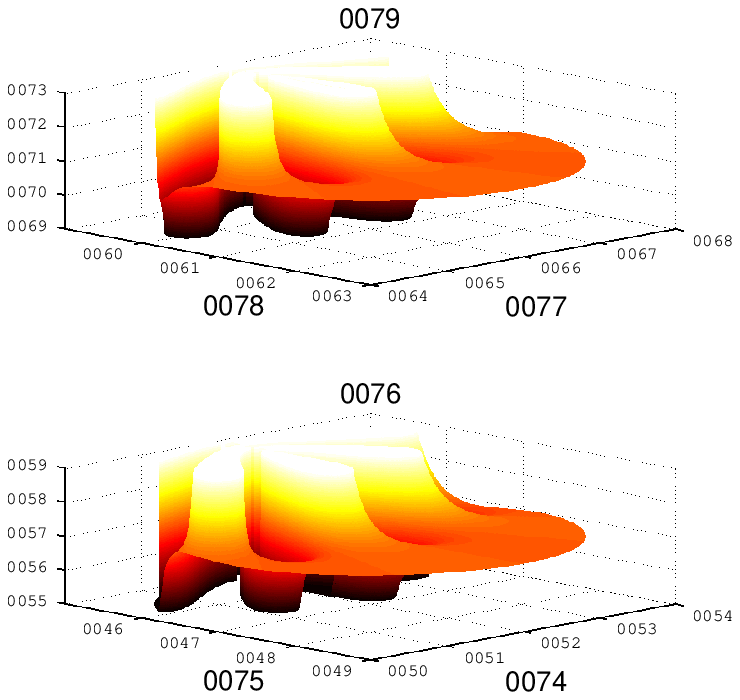} \label{g_complex}}
}
\caption{The Laplace transform $g(z)$ for  $f=0.1$ obtained by solving~\eqref{fe} iteratively.
%We see that $|g(z)|$ is a power-law at large $|z|$, whose logarithmic slope approaches
%$\alpha=-1/\log_2(1-0.1)=6.58$ as expected from~\eqref{alpha}.
\subref{gz0.1} 
$\mathrm{Re}[g(z)]$ (top left panel), $\mathrm{Im}[g(z)]$ (bottom left panel), and $|g(z)|$ (right panel)
versus $r$  along the real axis (dashed curve), the imaginary axis (solid curve),
and the line inclined at $\theta=60^{\circ}$ to the real axis (dotted curve). 
The variables $r$ and $\theta$ are defined by $z=r e^{i\theta}$.
\subref{g_complex}
A view of the real (top) and imaginary (bottom) parts of $g(z)$
(values above $1$ and below $-1$ have been cut off).}
\end{figure}
An example of the numerical solution for $f=0.1$ is presented in Figures~\ref{gz0.1} and~\ref{g_complex}.
The power-law behaviour at large $|z|$, with the exponent given by~\eqref{alpha}, is confirmed numerically
for $f=0.1$ (right panel of Figure~\ref{gz0.1}) and a range of other values. 
The iterations converge in the negative half-plane, $\mathrm{Re}(z)<0$,
despite the complicated structure of $g(z)$, illustrated in Figure~\ref{g_complex},
as it gradually approaches $e^{-z}$ for decreasing values of $f$.
The convergence does not depend on the initial function $g_0(z)$; 
e.g.\  $g_0(z)=e^{-z}$ works just as well for small values of $f$.

\section{Steady-state wealth distribution by Laplace inversion}
\label{section.laplace}
\begin{figure}[t!]
{\centering
\subfigure[linear scale]{\includegraphics{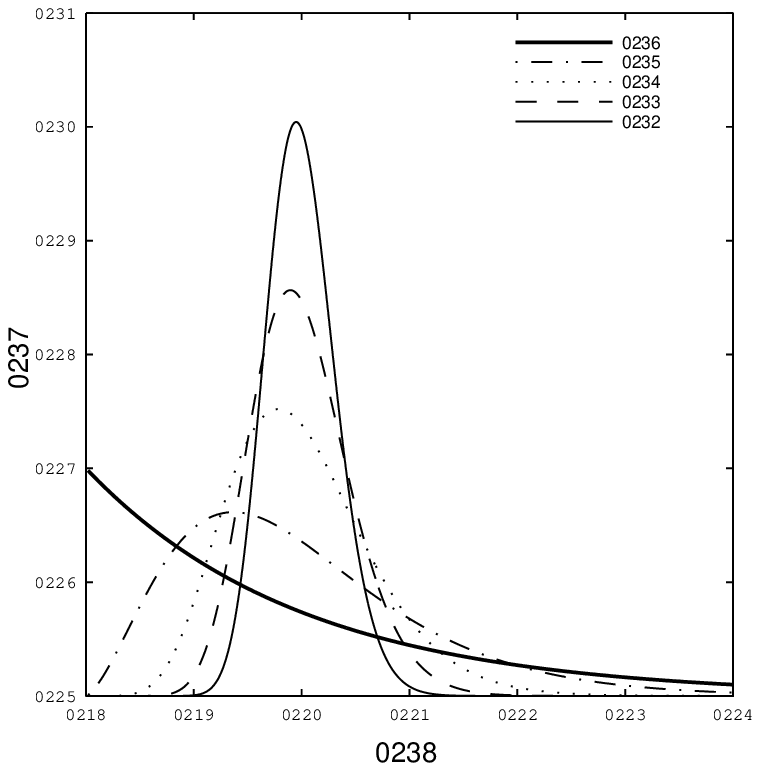} \label{inverse_Laplace_transforms}}
\subfigure[log-linear scale]{\includegraphics{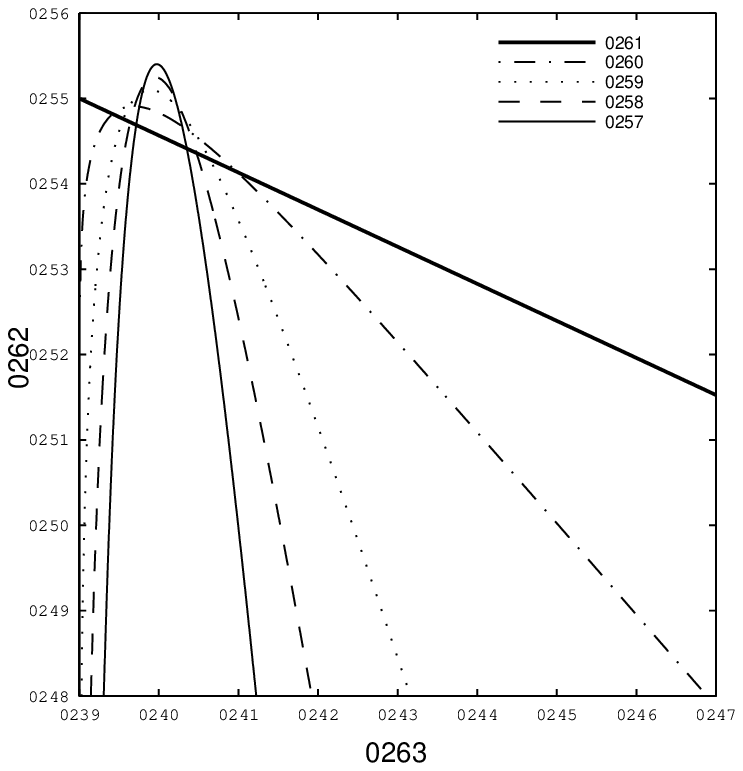} \label{inverse_Laplace_transforms_log}}
}
\caption{The steady-state wealth probability distribution function $p_s(w)$ 
obtained by inverting the Laplace transform $g(z)$ for the following values of the transfer fraction: 
$f=0.5$ (bold solid curve), $0.25$ (dash-dot curve), $0.1$ (dotted curve), $0.05$ (dashed curve), and $0.025$ (thin solid curve).
%$f=0.5$ (bold black curve), $0.25$ (cyan curve), $0.1$ (blue curve), $0.05$ (green curve), and $0.025$ (red curve).
}
\end{figure}

\begin{figure}[t!]
{\centering
\subfigure[log-log scale]{\includegraphics{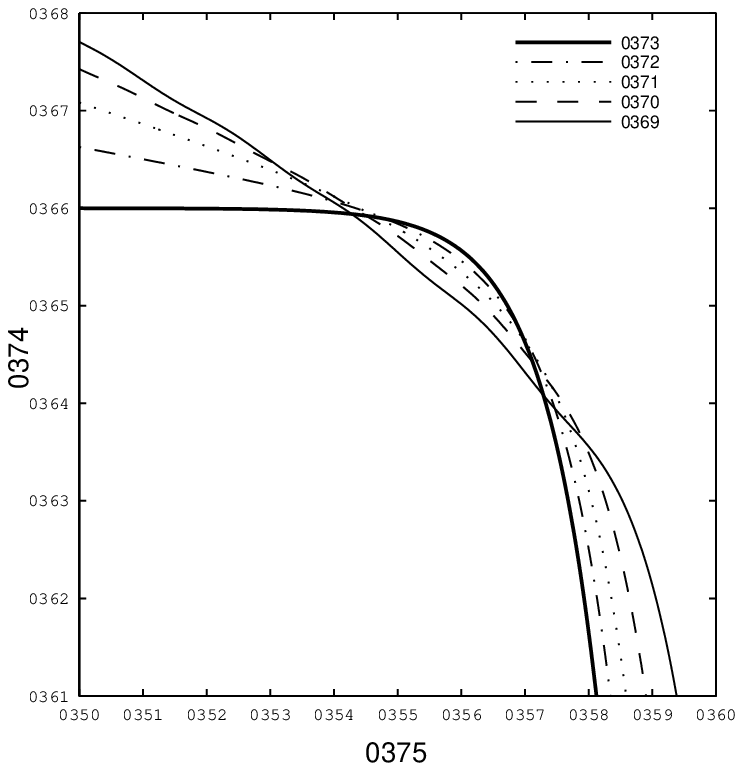}\label{inverse_Laplace_transforms_largef}}
\subfigure[log-linear scale]{\includegraphics{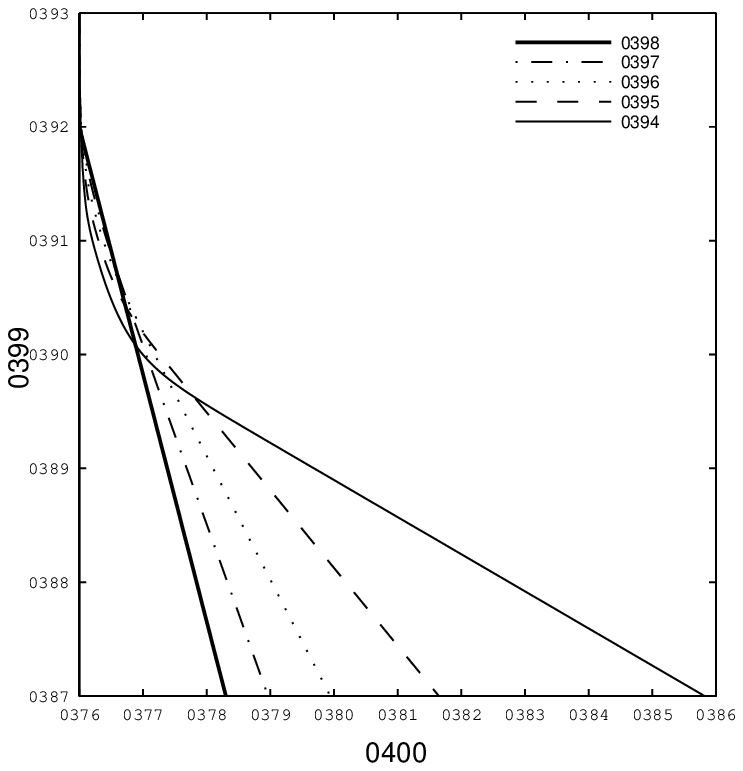}\label{inverse_Laplace_transforms_largef_semilog}}
}
\caption{The steady-state wealth probability distribution function $p_s(w)$ 
obtained by
inverting the Laplace transform $g(z)$ 
for the following values of the transfer fraction: 
$f=0.5$ (bold solid curve), $0.6$ (dash-dot curve), $0.7$ (dotted curve), $0.8$ (dashed curve), and $0.9$ (thin solid curve).
%$f=0.5$ (bold black curve), $0.6$ (cyan curve), $0.7$ (blue curve), $0.8$ (green curve), and $0.9$ (red curve).
}
\end{figure}

\begin{figure}[t!]
{\centering
\subfigure[$f=0.95$]{\includegraphics{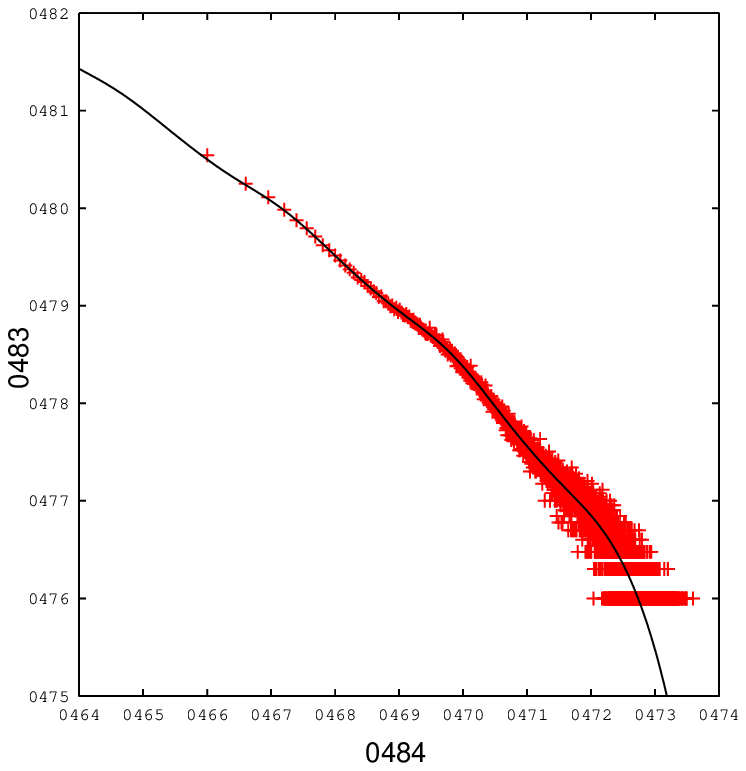} \label{compare_largef}}
\subfigure[$f=0.05$]{\includegraphics{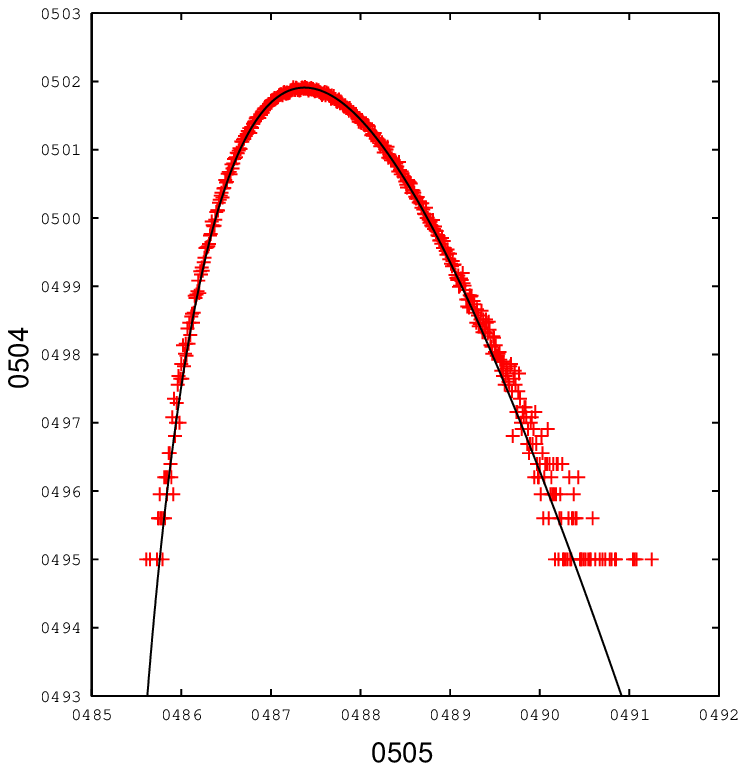} \label{compare_smallf}}
}
\caption{The population distribution $n(w)$, shown with crosses,
as a function of wealth $w$, measured in ficticious monetary units (m.u.) used in the agent-based simulations.
The distribution is computed as the number of agents in every unit wealth interval
after 100 steps in the simulation of the giver scheme with total number of agents $N=4\times10^5$ 
and transfer parameter \subref{compare_largef} $f=0.95$ and \subref{compare_smallf} $f=0.05$.
The initial distribution is uniform in the wealth interval 
\subref{compare_largef} $[0,100\,\mathrm{m.u.}]$ and \subref{compare_smallf} $[0,500\,\mathrm{m.u.}]$.
The corresponding solution of the steady-state master equation for the same $f$ 
is shown with a solid curve, with $p_s(w)$ scaled to conform with the
definition of $n(w)$ according to $Np_s(w/\langle w\rangle)/\langle w\rangle$ where $\langle w\rangle$ is the mean wealth.
Both the agent-based simulations and the master equation predict oscillations in the wealth distribution
in \subref{compare_largef} but not in \subref{compare_smallf}.}
\end{figure}

The steady-state probability distribution function $p_s(w)$ can be obtained by inverting its Laplace transform $g(z)$ numerically.
A number of inversion algorithms were reviewed recently in \cite{hassanzadeh2007comparison} and \cite{abate2006unified}.
The reviewers advised that at least two different algorithms should be used as a cross-check, 
because different algorithms work well for specific classes of functions and none of the algorithms is universally accurate.
Fortunately, the algorithms are easy to implement.
We test four 
(referred to as the Euler, Talbot, Stehfest, and Zakian algorithms in \cite{hassanzadeh2007comparison,abate2006unified}) and find
that the first two give accurate results over a wider range of $w$.
Euler has an additional advantage over Talbot: it samples $g(z)$ in the positive half-plane only,
where the function $g(z)$ has a simpler structure, as one sees in Figure~\ref{g_complex}.
The results of the inversion are presented in 
Figures~\ref{inverse_Laplace_transforms} and~\ref{inverse_Laplace_transforms_log} for $0.025\le f\le0.5$ 
and 
Figures~\ref{inverse_Laplace_transforms_largef} and~\ref{inverse_Laplace_transforms_largef_semilog} for $0.5\le f\le 0.9$.
The exponential analytic solution is recovered numerically for $f=0.5$.
From the output of the inversion algorithms, we compute the moments of the distribution ($\mu_0$, $\mu_1$, and $\mu_2$)
and find agreement with the analytical results to 8~significant digits.
In Figures~\ref{compare_largef} and~\ref{compare_smallf} for $f=0.95$ and $f=0.05$ respectively, we compare the wealth distributions
obtained from the Laplace transform (curves) and from the agent-based simulations (crosses).
We find excellent agreement between these two methods for all values of $f$ that we consider.

The algorithms that perform the inverse Laplace transform suffer from truncation errors.
Maximum precision is achieved near the peak of $p_s(w)$.
In the tail,
the precision decreases until the results are completely dominated by the truncation errors
below a threshold value of $p_s(w)$.
For example,
we perform all computations with 16 significant digits and achieve
$\sim8$ significant digits of precision at the peak of $p_s(w)$,
but
the algorithms break down at $p_s(w)\lesssim10^{-8}$.

The wealth distributions that we find for $f<1/2$ are characterized by 
power-law behaviour at $w\ll1$, in accord with the analytical results, and approximately exponential tails at large $w$.
A careful examination of the tails confirms that the asymptotic behaviour at large wealths is not exactly exponential.
However, we have not been able to find a closed-form expression for it.
The distribution becomes tightly concentrated around its peak as $f$ decreases;
the peak of the distribution gradually shifts towards $w=1$.
On the other hand, the peak shifts towards $w=0$ as $f$ increases;
the distributions eventually turns into an exponential function for $f=1/2$.
This overall behaviour is similar to 
that observed in the asset exchange models based on the saving propensity with $0<\lambda<1$ \cite{chatterjee2007kinetic}.

The structure of the steady-state solutions for $f>1/2$ is very different.
The asymptotic approximation $p_s(w)\propto w^{\alpha-1}$, with $\alpha=-1/\log_2(1-f)$, is valid for $f>1/2$ as well
and indicates that
the distribution diverges at $w=0$ (as $\alpha<1$).
For values of $f$ sufficiently close to $1$,
the wealth distribution acquires a shape that is akin to a power law, 
$p_s(w)\propto w^{-1}$,
with overlaid \textit{oscillations} that become more prominent as $f$ increases.
This power-law behaviour cuts off exponentially at some critical wealth that increases slowly as $f$ approaches $1$.
At the same time the exponential drop-off at large wealths becomes shallower 
as evident from Figure~\ref{inverse_Laplace_transforms_largef_semilog}.
The oscillations of $p_s(w)$ appear to be periodic on a logarithmic scale, with the period depending on $f$.
For example, the periods for $f=0.9$ and $f=0.99$ are roughly one and two decades respectively.
This is directly related to the fact that all givers retain 1\% of their wealth for $f=0.99$
and 10\% for $f=0.9$.

\section{Discussion}
\label{section.discussion}
We now use the Laplace transform tools developed in Section~\ref{section.laplace}
to address two questions that are costly to explore with agent-based simulations:
the nature of disorder (entropy) and its evolution in the giver scheme,
and the degree of inequality in the steady state.

\subsection{Entropy and the approach to equilibrium}
According to Boltzmann, states with higher entropy are more probable 
because they correspond to a larger number of microscopic configurations of the system.
A closed system evolves to a state of maximum entropy, i.e.\ maximum disorder,
which is characterized by the Boltzmann-Gibbs distribution.
The exponential distribution observed in models 
where the transfer amount is fixed and constant (see section II.C in \cite{yakovenko2009colloquium})
is thus consistent with entropy maximization ideas.
On the other hand, it is argued in \cite{yakovenko2009colloquium} that
multiplicative asset exchanges may lead to non-exponential distributions because of the broken time-reversal symmetry,
whereas, in models with fixed additive exchanges, the time-reversal symmetry is preserved.
Despite this, the entropy maximization technique has been applied in \cite{chakraborti2008gamma}
to an asset exchange model described by a Hamiltonian quadratic in wealth variables. 
It predicts a gamma distribution of wealth,
but $p_s(w)$ in multiplicative asset exchange models is not a gamma distribution in general.
\begin{figure}[t!]
{\centering
\subfigure[steady-state entropy]{\includegraphics{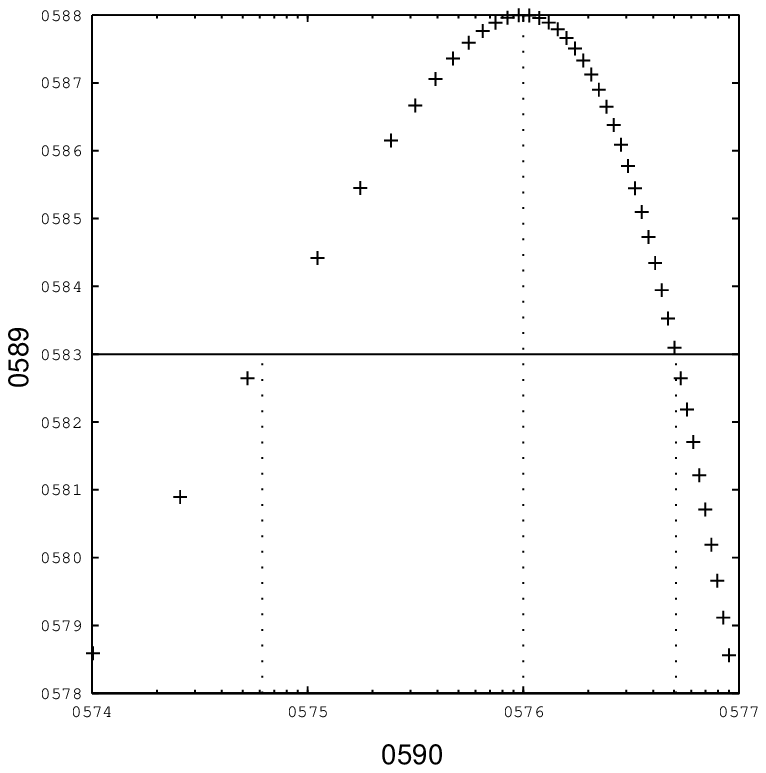}\label{entropy}}
\subfigure[entropy evolution]{\includegraphics{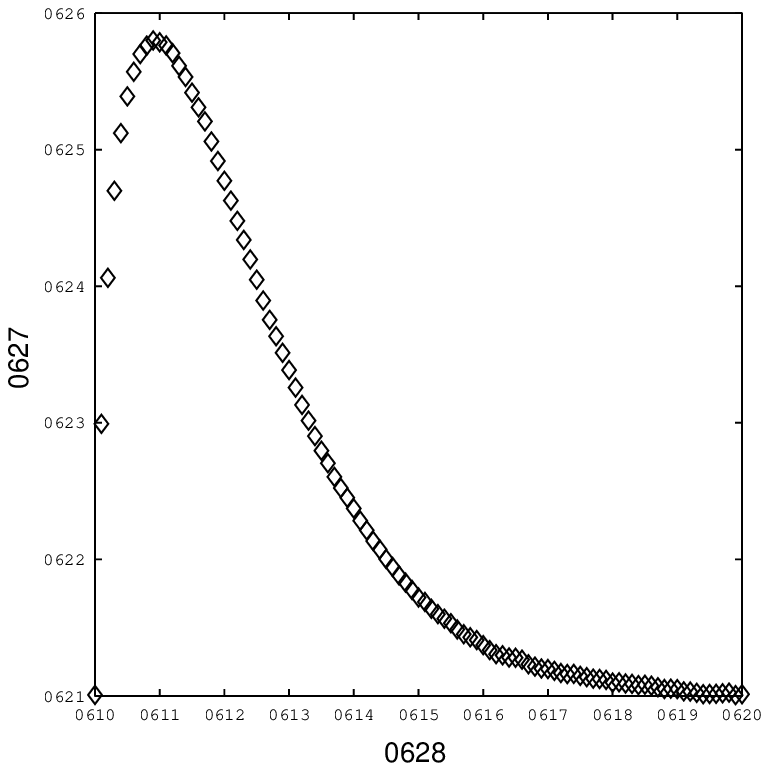}\label{entropy_evolution}}
}
\caption{\subref{entropy} Boltzmann entropy $S_s$ of the steady-state distribution 
as a function of the variance $\sigma_s^2=f/(1-f)$.
The critical values $\sigma_s^2=0.062$, $1$, and $5.098$, corresponding to $f=0.058$, $0.5$, and $0.836$ respectively, 
are indicated with the dotted lines.
\subref{entropy_evolution} Entropy as a function of time for the initial distribution given by~\eqref{initial_zero_entropy}
with $f=0.058$, computed from the multi-agent simulation of the giver scheme.
For the simulation, the distribution~\eqref{initial_zero_entropy} was scaled up to give $N=337123$~agents in $0\le w\le1421$.
To compute the entropy, the population distribution produced by the simulation was normalized to a probability distribution with unit mean.
}
\end{figure}

In this section, we explore the applicability of the Boltzmann entropy to the giver scheme.
Using the Laplace transformed solutions of the master equation~\eqref{master}, 
we compute the steady-state Boltzmann entropy\footnote{We stress 
that this definition applies to a normalized distribution with unit mean.
Unlike the case of discrete probability distributions, continuous entropy can be negative and it is not 
invariant with respect to the change of variable.
It may be more appropriate to consider the Kullback-Leibner divergence $D_s=\int_{0}^{\infty}dw\;p_s(w)\log[p_s(w)/m(w)]$, which
is a measure of the divergence between $p_s(w)$ and the reference distribution $m(w)$.
It is convenient to take $m(w)=e^{-w}$, in which case the divergence essentially reverts to Boltzmann entropy
because $D_s=1-S_s$.}
\begin{equation}\label{boltzmann}
S_s=-\int_{0}^{\infty}
dw\;
p_s(w)\log[p_s(w)]
\end{equation}
for several values of $f$ in the range $0.01\le f\le0.9$ and plot the results in Figure~\ref{entropy}.
Entropy maximization arguments~\cite{kapur1989maximum} imply that the entropy defined by~\eqref{boltzmann}
leads uniquely to the exponential distribution, $p(w)=e^{-w}$ with $S_s=1$, 
if the only condition is that the mean of the distribution is fixed to $\mu_1=1$.
In our model, however, the transfer fraction $f$ places additional constraints on how
the distribution of wealth evolves with time.
Therefore, it is not surprising that we observe a range of steady-state entropies $S_s\ne1$
corresponding to different values of $f$.
The exponential distribution for $f=1/2$ appears to be the most disordered state of the system
with the highest entropy $S_s=1$.
For all other values of $f$, the entropy is smaller and it becomes negative as $f$ approaches $0$ or $1$.
Reading off the graph, we find that $S_s$ is negative for $0<f<0.058$ and $0.836<f<1$.
The entropy tends to negative infinity as $f\to0$ or $f\to1$, which is in accord with 
the behaviour of $p_s(w)$ in these limits.

Negative values of $S_s$ are already a warning that the Boltzmann entropy may not be a faithful measure of disorder
in a multiplicative asset transfer system like the giver scheme.
However, the situation worsens when we look at how $S(t)=-\int_{0}^{\infty}dw\;p(w,t)\log[p(w,t)]$ evolves with time
by conducting multi-agent simulations.
In many cases, it decreases instead of increasing.
For example, if we choose $p(w,0)=e^{-w}$ initially, $S(t)$ decreases with time for all $f\ne1/2$
and remains constant for $f=1/2$.
Moreover, we can easily find realistic situations where $S(t)$ does not change monotonically with time,
as the experiment described below shows.
Consider an initial distribution of the form
\begin{equation}\label{initial_zero_entropy}
p(w,0)
=
\left\{
\begin{array}{cl}
p_1, \quad &0\le w\le 1,\\
p_2, \quad &1<w\le w_2,\\
0,  \quad &\textrm{otherwise},
\end{array}
\right.
\end{equation}
with the parameters $p_1$, $p_2$, and $w_2$ chosen to give $S(0)=0$ ($p_1\approx0.296$, $p_2\approx1.669$, and $w_2\approx1.421$).
The evolution of entropy for this initial distribution with $f=0.058$ is plotted in Figure~\ref{entropy_evolution}.
The entropy grows initially but after about ten steps in the simulation it turns over and begins to decrease, 
eventually reaching $S_s=0$ as expected for $f=0.058$.
This is in marked contrast to the behaviour of $S(t)$ in an ideal gas, 
where one has $dS(t)/dt\ge0$ according to Boltzmann's H-theorem.

The population distribution is determined by dividing the wealth axis into small bins
and computing the number of agents that fall in each bin.
One can define the multiplicity $W$ as
the number of permutations of the agents between different wealth bins  such that the occupation numbers of
the bins do not change. 
The definition of entropy, $S_s=\log W$, leads to the expression~\eqref{boltzmann} in the continuous limit.
Entropy maximization under the assumption that total wealth is conserved gives an exponential distribution.
However, this ignores the global constraints on the probability distribution $p_s(w)$ imposed by the transfer fraction $f$.
The maximization procedure must take these constraints into account
to derive the steady-state wealth distribution appropriate to the giver scheme.
Unfortunately, at the time of writing, we have been unable to derive these additional constraints,
and they do not appear in the literature.
\begin{figure}[t!]
{\centering
\subfigure[]{\includegraphics{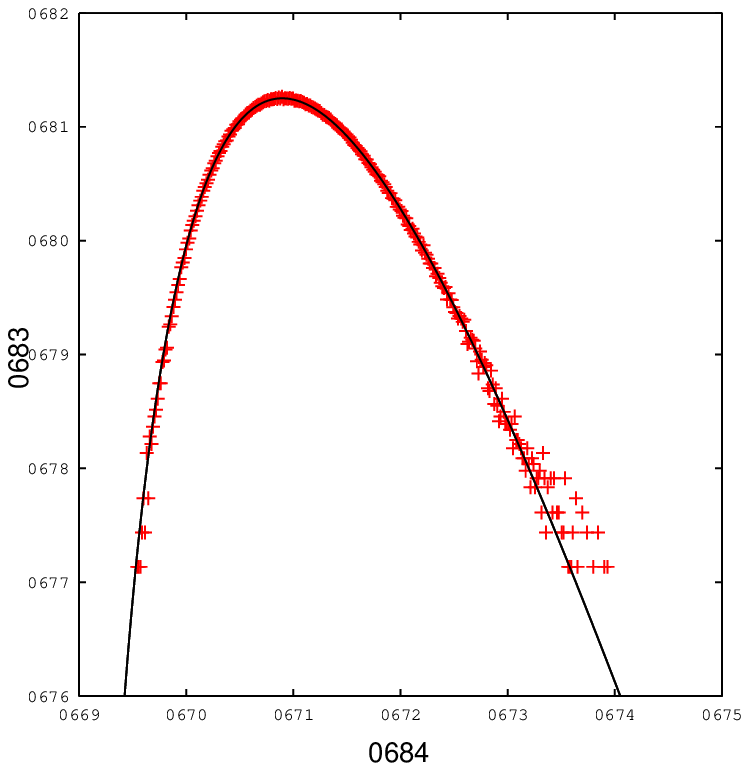}\label{random_process_distribution}}
\subfigure[]{\includegraphics{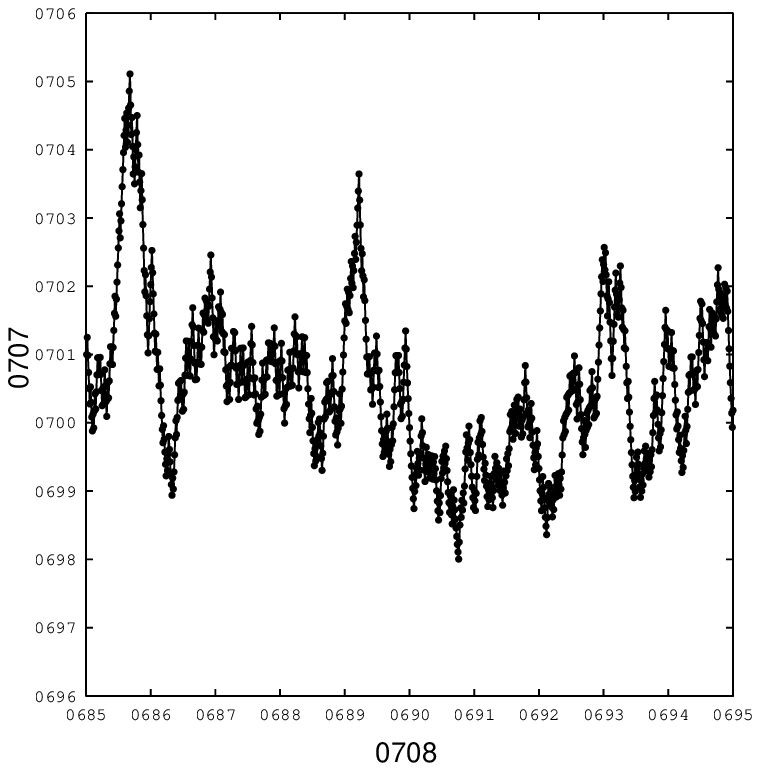}\label{random_process_view}}
}
\caption{\subref{random_process_distribution} The limiting probability distribution $\xi(w)$, shown with crosses, 
obtained from one realization of the asymmetric random process~\eqref{random_process_definition}
for $f=0.05$ after $10^6$ iterations.
The mean is 0.9998 and the variance is 0.0519 (cf.~0.0526 theoretically from the master equation).
The steady-state distribution $p_s(w)$ for the same $f$ obtained by Laplace inversion is shown as a solid curve for comparison.
\subref{random_process_view} The first 1000 values of $\{w_i\}$.}\label{random_process}
\end{figure}

\subsection{Random process}
The evolution of wealth in the giver scheme can be analyzed in terms of 
a random process defined by 
\begin{equation}\label{random_process_definition}
w_{i+1}=w_i+\Delta w_i,
\end{equation}
 where $w_1=1$ and $\Delta w_i=+f$ or $\Delta w_i=-fw_i$ with equal probability.
This process is asymmetric, 
i.e.\ multiplicative in the negative direction and additive in the positive.
To illuminate the relationship between the giver scheme and the random process we note that
(1) an agent's loss of wealth is always proportional to his wealth, i.e.~it is multiplicative, 
and (2) an agent's gain of wealth can originate from any other agent in the population
and therefore equals $f\langle w\rangle$ on average, or simply $f$ if we set $\langle w\rangle=1$.
We compute the limiting distribution~$\xi(w)$ of this process by applying~\eqref{random_process_definition} 
a sufficiently large number of times 
and then constructing a histogram of all $\{w_i\}$.
By the ergodic assumption, this is equivalent to computing a large number of realizations of this random process
and using the final values in each realization to find the limiting distribution.

Figures~\ref{random_process_distribution} and~\ref{random_process_view} 
display one particular realization of the random process~\eqref{random_process_definition}
and the corresponding limiting distribution.
Given the close link between the transfer model and the random process, it is not surprising that
the limiting distribution $\xi(w)$ of the random process~\eqref{random_process_definition}
appears to be identical to the steady-state distribution $p_s(w)$ of the giver scheme.
We obtain similar results for other values of $f\ll1$.
Note that the slight discrepancy between $\xi(w)$ and $p_s(w)$ at $w>2$ is due to insufficient sampling of large wealths by the random process.
The agreement improves as the number of iterations increases.

\subsection{Inequality of wealth}
A traditional measure of inequality in economic systems is the Gini coefficient,
defined as $G=1-2\int_{0}^{1}L(X)dX$, 
where $L(X)$ is the Lorenz curve. 
For a continuous distribution, we have $L(w)=\int_{0}^{w}dw'\; w'p_s(w')$, $X(w)=\int_{0}^{w}dw'\; p_s(w')$,\footnote{
Note that we have $L(X)\le X$ for all $X$ because $L(0)=0$, $L(1)=1$, and $dL/dX$ is a monotonically increasing function.}
and hence
\begin{equation}
G_s =
1-2
\int_{0}^{\infty}
dw\; p_s(w)
\int_{0}^{w}
dw'\; w'p_s(w')
,
\end{equation}
such that $G_s=0$ corresponds to perfect equality, and $G_s=1$ to perfect inequality.

\begin{figure}
\centering
\includegraphics{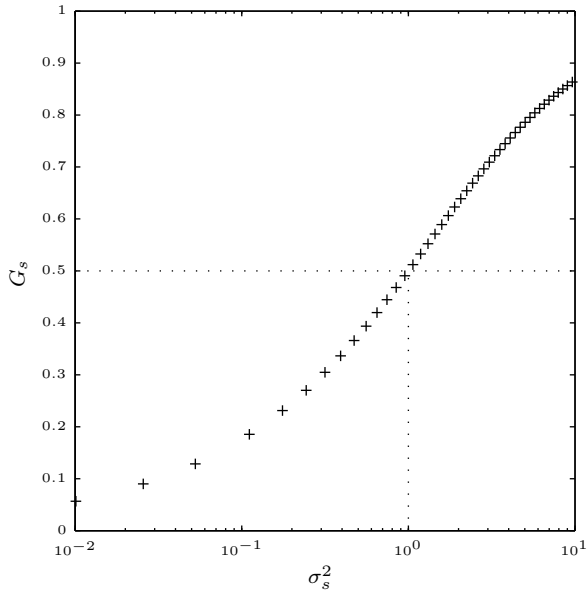}
\caption{Gini coefficient 
of the steady-state distribution $p_s(w)$
as a function of the variance $\sigma_s^2=f/(1-f)$.
}\label{gini}
\end{figure}

In Figure~\ref{gini} we plot $G_s$ versus the steady-state variance $\sigma_s^2=f/(1-f)$ for $0.01\le f \le 0.9$. 
As expected, $G_s$ increases monotonically with $\sigma_s^2$, since both quantities are measures of dispersion.
Interestingly, however, there is an inflection point in the $G_s(\sigma_s^2)$ curve at $\sigma_s^2=1$, $G_s=1/2$,
corresponding to the exponential distribution (i.e.~$f=1/2$).
For $f\to0$, we have $p_s(w)\to\delta(w-1)$,
which corresponds to perfect equality since all agents have the same wealth.
On the other hand, for $f\to1$, 
the distribution $p_s(w)$ becomes sharply peaked near~$w=0$, while the standard deviation approaches infinity.
This corresponds to the situation where most agents have zero wealth, except for one  who has everything, i.e.\ perfect inequality.

We can understand the evolution towards inequality in terms of the state vector of the system.
Consider  $N$ agents whose wealths are characterized by random variables $w_i$, $i=1,2,\dots,N$.
The state of the system can be described by the phase-space vector $\mathbf{w}=(w_1,w_2,\dots,w_N)$.
The constraints that define the phase-space are 
(1) $0\le w_i\le1$ for all $i$, and
(2) $\sum_{i=1}^{N}w_i=1$ (we assume for convenience that the total wealth is unity).
These constraints define a segment of the $(N-1)$-dimensional hyperplane embedded in $N$-dimensional space.
Without any additional constraints, entropy maximization \cite{kapur1989maximum} gives $g_i(w_i)=(N-1)(1-w_i)^{N-2}$
for the probability distribution of the wealth $w_i$ of the $i$-th agent,
with mean $\langle w_i\rangle = \frac{1}{N}$ and variance $\sigma^2_{w_i}=\frac{N-1}{N^2(N+1)}$.
However, in our system, the asset transfer process and the value of the parameter $f$
place additional restrictions on the evolution of $\mathbf{w}$.
For the increment of the phase-space vector $\mathbf{w}$ from time $t_k$ to time $t_{k+1}$, i.e.\ after one generation of asset transfers,
we have
\begin{equation}
\norm{\Delta\mathbf{w}}^2
= 
\sum_{i=1}^{N}[w_i(t_{k+1})-w_i(t_k)]^2.
\end{equation}
%The change $\Delta\mathbf{w}$ of $\mathbf{w}$ from time $t_k$ to time $t_{k+1}$, i.e.\ after one generation of asset transfers,
%is given by 
%\begin{equation}
%|\Delta\mathbf{w}|^2
%= 
%\sum_{i=1}^{N}[w_i(t_{k+1})-w_i(t_k)]^2.
%\end{equation}
The terms in the sum on the right hand side can be split into two groups, associated with the givers and the receivers.
Since the transfer amount is proportional to the giver's wealth, we get
\begin{equation}
\norm{\Delta\mathbf{w}}^2
= 
2f^2\sum_{i\in\text{givers}}[w_i(t_k)]^2 
.
\end{equation}
Therefore the following inequality must always be satisfied:
\begin{equation}
\norm{\Delta\mathbf{w}}
\le
2^{1/2}f\norm{\mathbf{w}}
.
\end{equation}
In addition, we have
\begin{equation}
N^{-1/2}\le \norm{\mathbf{w}}\le1
\end{equation}
due to the restrictions of the phase-space itself. 
Note that the state $w_i=1/N$ for all $i$ is the nearest point to the origin.

When $f$ is small, the norm of the increment $\norm{\Delta\mathbf{w}}$
is also small compared with the maximum linear extent of the phase space (which equals $2^{1/2}$);
 the evolution of $\mathbf{w}$ is gradual.
Furthermore, $\norm{\Delta\mathbf{w}}$ is also constrained by $\norm{\mathbf{w}}$,
which can be very small if $N$ is large and 
all agents are clustered as close to the origin as possible.
So, if the dispersion in wealth is modest,
$\mathbf{w}$ moves slower through the phase space than if there is great inequality.
On the other hand, $\mathbf{w}$ changes more rapidly on average if $\norm{\mathbf{w}}$ is close to unity,
which corresponds to large inequality.
The states of equality are therefore more probable, which explains 
why the steady-state distribution tends towards a delta function for $f\to0$.
Even if the initial wealth distribution is very unequal, $\mathbf{w}$ drifts quickly towards
the states of near equality.

When $f$ is close to 1, $\norm{\Delta\mathbf{w}}$ can be comparable to the size of the phase space.
Since $f$ is large, the gains in wealth of the individual agents can be large as well.
This leads to the situation where a few agents own most of the wealth.
These agents retain their large wealth for a short time only (typically a few time steps)
before they become givers and pass their large wealth to other agents.
In the extreme case $f=1$, one agent possesses all the wealth at any instant,
while all the other agents have zero wealth.
This maximum wealth is passed from agent to agent frequently.
This corresponds to $\mathbf{w}$ jumping from one corner of the phase space to another.
For $f\approx1$, $\mathbf{w}$ evolves similarly, with $p_s(w)$ 
peaking strongly at zero wealth.

\section{Conclusions}
\label{section.conclusions}

We develop a new technique for computing the steady-state probability distribution of 
a multiplicative asset transfer model, which we call the giver scheme, by Laplace transforming the associated master equation
to give a functional equation for the characteristic function of the distribution.
In the giver scheme, the transfer amount $fw_g$ is proportional to the giving agent's wealth $w_g$,
so the model depends on a single parameter $f\in(0,1)$.
We develop an efficient iterative method to solve the functional equation for any $f$,
and we employ several Laplace inversion algorithms to recover the steady-state distribution $p_s(w)$.

We comprehensively explore the dependence of the wealth distribution on the value of $f$,
 especially the thinly studied regime $1/2\le f<1$.
We find a stark qualitative difference between the distributions 
 for $f\approx0$ (sharply peaked distribution centred around the mean wealth) 
 and $f\approx1$ (broad distribution of approximately power-law shape with overlaid oscillations). 
These two extremes correspond to near-perfect equality and inequality respectively,
 as characterized by the Gini coefficient.
Both extremes are also characterized by negative Boltzmann entropy.
While the regime $f\approx0$ is 
 generally thought to represent to some extent the exchange processes occuring in the real economy,
 the regime $f\approx1$ is probably less applicable to realistic economic systems,
 except perhaps in situations involving extreme leverage.
The regime $f\approx1$ may also be relevant to the analysis of gambling,
 where transitory fortunes are made and lost frequently.

We show that the Boltzmann entropy is unlikely to be a faithful measure of disorder in a multiplicative asset transfer system,
since it does not vary monotonically as a function of time, assuming the second law of thermodynamics.
This is an important and counterintuitive result, because the system in the giver scheme is closed and the microscopic transfer rules conserve wealth,
in a manner reminiscent of the microcanonical ensemble in statistical mechanics.
%It may be fruitful to consider more general definitions of entropy that
%allow for a parametric dependence on $f$, such as the Tsallis entropy (see also \cite{kaniadakis2009maximum}),
%at least for power-law distributions.
In a multiplicative transfer system, 
the correlations between various subsystems 
(e.g.\ subclasses corresponding to a particular historical sequence of giving and receiving)
and the time-reversal asymmetry of the microscopic rules
are crucial to the system's dynamics and, therefore, cannot be ignored.

\appendix

\section{Iterative procedure}
\label{appendix.iterations}
We assume that the computations are carried out with 16 significant digits.
For a given complex argument $z$, define a uniform grid $u=\{u_k\}_1^K$ that covers 
the interval $[-4,\log_{10}(|z|)]$. 
The approximation~\eqref{g0} gives sufficient precision for $|z|<10^{-4}$
for computations with 16 significant digits.
Choose the number of points $K$ such that there
are a large number of points in every unit interval, say, $10^3$ logarithmic grid points per decade in $[10^{-4},|z|]$.
Define two auxiliary grids, $u_{(f)}=\log_{10}(f)+u$ and $u_{(1-f)}=\log_{10}(1-f)+u$,
and initialize the iterations with $g_0(10^u z/|z|)=1/(1+10^u z/|z|)$.
For a given set of values $g_i(10^u z/|z|)$, defined on the grid $u$,
find the corresponding values on the auxiliary grids by performing a spline interpolation or 
using the approximation~\eqref{g0} where appropriate.
Then use these values in equation~\eqref{iteration equation} to find $g_{i+1}(10^u z/|z|)$.
Continue iterating until the convergence criterion is met
(we find that the convergence spreads gradually from zero to $|z|$).
Typically the convergence requires a few dozens of iterations for $|z|\sim100$ in the positive half-plane.
Once the convergence is reached, apply a spline interpolation to find $g(z')$ for any $z'$
along the same direction in the complex plane as $z$, provided that $|z'|<|z|$.

The obvious disadvantage of the procedure outlined above is that it relies on interpolation.
Its precision is therefore limited by the number of points in the grid $u$, 
i.e.\ the discretization of the interval $[0,|z|]$.
It is possible, however, to avoid interpolation altogether by defining a special non-uniform grid
that is invariant with respect to division by $f$ and $(1-f)$.
This gives rise to an alternative procedure for computing the iterations.

Define a grid $r_{k,m}=f^k(1-f)^m$ with $0\le k\le K$ and $0\le m\le M$, where
$K=\lceil \log(10^{-4}/|z|)/\log(f) \rceil$ and $M=\lceil \log(10^{-4}/|z|)/\log(1-f) \rceil$
are defined such that $|z|r_{K,0}<10^{-4}$ and $|z|r_{0,M}<10^{-4}$.
The function $g(z)$ on the grid $z_{k,m}=r_{k,m}z$, defined according to $g_{k,m}=g(r_{k,m}z)$,
has the following properties:
$g(fz_{k,m})=g(z_{k+1,m})=g_{k+1,m}$ and $g[(1-f)z_{k,m}]=g(z_{k,m+1})=g_{k,m+1}$.
Therefore the iteration rule becomes
\begin{equation}
g_{k,m}=\frac{g_{k,m+1}}{2-g_{k+1,m}},
\end{equation}
for $0\le k \le K-1$ and $0\le m \le M-1$.
For $g_{K,M}$ one can use the approximation \eqref{g0}.
In fact, the Taylor expansion can be used for any point $|z_{k,m}|<10^{-4}$.
Thus, no interpolation is required and the iterations can be computed more efficiently. 
However, unlike the approach based on interpolation, this procedure must be 
repeated for different arguments even if they lie 
in the same direction in the complex plane.

\section*{Acknowledgement}
AS acknowledges generous financial support from the Portland House Foundation.

\section*{References}
\bibliographystyle{elsarticle-num}
\bibliography{asokolov.bib}

\end{document}